\documentclass[12pt,draftcls,onecolumn,journal]{IEEEtran}
\usepackage{amsfonts}
\usepackage{amsfonts}
\usepackage{setspace}

\usepackage{amsmath}
\usepackage{amssymb}
\usepackage[dvips]{graphicx}
\usepackage{setspace}
\usepackage{epsfig}
\newcommand{\K}{{\mathbf{K}}}
\newcommand{\tK}{{\widetilde{\mathbf{K}}}}
\newcommand{\x}{\mathbf{x}}
\newcommand{\y}{\mathbf{y}}

\newcommand{\h}{\mathbf{h}}
\newcommand{\n}{\mathbf{n}}
\newcommand{\hh}{\mathbf{H}}

\newcommand{\I}{\mathbf{I}}

\newcommand{\E}{\mathbb{E}}

\newcommand{\bv}{\mathbf{v}}
\newcommand{\dd}{\dagger}

\newcommand{\tr}{{\text{tr\,}}}
\newcommand{\0}{\mathbf{0}}
\newcommand{\bu}{\mathbf{u}}
\newcommand{\U}{\mathbf{U}}
\newcommand{\bLambda}{\mathbf{\Lambda}}
\newcommand{\bPhi}{\mathbf{\Phi}}
\newcommand{\lm}{\lambda_{\max}}

\newcommand{\dc}{\dot{C}_s(0)}
\newcommand{\ddc}{\ddot{C}_s(0)}
\newcommand{\so}{\mathcal{S}_0}

\newcommand{\figsize}{0.7}

\newcommand{\tsnr}{{\text{\footnotesize{SNR}}}}

\newtheorem{theo}{Theorem}
\newtheorem{prop}{Proposition}
\newtheorem{rem}{Remark}
\newtheorem{cor}{Corollary}

\begin{document}

\title{Secure Communication in the Low-SNR Regime
}



%

\author{\vspace{.5cm}\authorblockN{Mustafa Cenk Gursoy}
\thanks{The author is with the Department of Electrical
Engineering, University of Nebraska-Lincoln, Lincoln, NE, 68588
(e-mail: gursoy@engr.unl.edu).}
\thanks{The material in this paper was presented
in part at the IEEE International Symposium on Information Theory (ISIT), in Seoul, Korea in June 2009.}
\thanks{This work was supported by the National Science Foundation under CAREER Grant CCF -- 0546384.}}


\maketitle

\begin{abstract} Secrecy capacity of a multiple-antenna wiretap channel is studied in the low signal-to-noise ratio ($\tsnr$) regime. Expressions for the first and second derivatives of the secrecy capacity with respect to $\tsnr$ at $\tsnr = 0$ are derived. Transmission strategies required to achieve these derivatives are identified. In particular, it is shown that it is optimal in the low-$\tsnr$ regime to transmit in the maximal-eigenvalue eigenspace of $\bPhi = \hh_m^\dd \hh_m - \frac{N_m}{N_e}\hh_e^\dd \hh_e$ where $\hh_m$ and $\hh_e$ denote the channel matrices associated with the legitimate receiver and eavesdropper, respectively, and $N_m$ and $N_e$ are the noise variances at the receiver and eavesdropper, respectively. Energy efficiency is analyzed by finding the minimum bit energy required for secure and reliable communications, and the wideband slope. Increased bit energy requirements under secrecy constraints are quantified. Finally, the impact of fading is investigated, and the benefits of fading in terms of energy efficiency are shown.


\end{abstract}

%

\begin{spacing}{1.7}
\section{Introduction} \label{sec:intro}

Secure transmission of confidential messages is a critical issue in communication systems and especially in wireless systems due to the broadcast nature of wireless transmissions. In \cite{Wyner}, Wyner addressed the transmission security from an information-theoretic point of view, and identified the rate-equivocation region and established the secrecy capacity of the discrete memoryless wiretap channel in which the wiretapper receives a degraded version of the signal observed by the legitimate receiver. The secrecy capacity is defined as the maximum communication rate from the transmitter to the legitimate receiver, which can be achieved while keeping the eavesdropper completely ignorant of the transmitted messages. Later, these results are extended to Gaussian wiretap channel in \cite{Hellman}. In \cite{Csiszar}, Csisz\'ar and K\"{o}rner considered a more general wiretap channel model and established the secrecy capacity when the transmitter has a common message for two receivers and a confidential message to only one. Recently, there has been a flurry of activity in the area of information-theoretic security, where, for instance, the impact of fading, cooperation, and interference on secrecy are studied (see e.g., \cite{specialissue} and the articles and references therein). Several recent results also addressed the secrecy capacity when multiple-antennas are employed by the transmitter, receiver, and the eavedropper \cite{Hero}--\cite{Oggier}. The secrecy capacity for the most general case in which arbitrary number of antennas are present at each terminal has been established in \cite{Khisti} and \cite{Oggier}.

In addition to security issues, another pivotal concern in most wireless systems is energy-efficient operation especially when wireless units are powered by batteries. From an information-theoretic perspective, energy efficiency can be
measured by the energy required to send one information bit
reliably. It is well-known that for unfaded and fading Gaussian channels subject to
average input power constraints, energy efficiency improves as one operates at lower $\tsnr$ levels, and the minimum bit energy is achieved as  $\tsnr$ vanishes \cite{Verdu}. Hence, requirements on energy efficiency necessitate operation in the low-$\tsnr$ regime. Additionally, operating at low $\tsnr$ levels has its benefits in terms of limiting the interference in wireless systems.

In this paper, in order to address the two critical issues of security and energy-efficiency jointly, we study the secrecy capacity in the low-$\tsnr$ regime. It is worthwhile to note that operation at low $\tsnr$s, in addition to improving the energy efficiency, is beneficial from a security perspective as well. In the low-$\tsnr$ regime, either the transmission power is small or the bandwidth is large. In either case, we have low probability of intercept as it is generally difficult for an eavesdropper to detect the signals in this regime.

We consider a general multiple-input and multiple-output (MIMO) channel model and identify the optimal transmission strategies in the low-$\tsnr$ regime under secrecy constraints. Since secrecy capacity is in general smaller than the capacity attained in the absence of confidentiality concerns, energy per bit requirements increase due to secrecy constraints. In this work, we quantify these increased energy costs and address the tradeoff between secrecy and energy efficiency. The main contributions of the paper are listed below:
\begin{enumerate}
\item We determine the first and second derivatives of the secrecy capacity at $\tsnr = 0$, and provide a second-order approximation to the MIMO secrecy capacity in the low-$\tsnr$ regime. Through this analysis, we quantify the impact of secrecy constraints on the performance.

\item We identify the optimal transmission strategies in the low-$\tsnr$ regime. In particular, we determine that transmission in the maximal-eigenvalue eigenspace of a certain matrix that depends on the channel matrices is second-order optimal. In the case in which the maximum eigenvalue is distinct, beamforming is shown to be optimal.

\item We find the minimum energy required to send one bit both reliably and securely. We characterize the tradeoff between energy efficiency and secrecy.

\item We investigate the impact of fading by studying the low-$\tsnr$ secrecy capacity in fading scenarios. We show that in general both independent and correlated fading improves the energy efficiency.

\end{enumerate}

The remainder of the paper is organized as follows. In Section \ref{sec:channelmodel}, we describe the channel model. In Section \ref{sec:lowsnrsecrecy}, we study the secrecy capacity in the low-$\tsnr$ regime and determine the minimum energy per secret bit. We investigate the impact of fading in Section \ref{sec:fading} and provide conclusions in Section \ref{sec:conclusion}. Lengthy proofs are relegated to the Appendix.

\section{Channel Model} \label{sec:channelmodel}

We consider a MIMO channel model and assume that the transmitter, legitimate receiver, and eavesdropper are equipped with $n_T, n_R$, and $n_E$ antennas, respectively. We further assume that the channel input-output relations between the transmitter and legitimate receiver, and the transmitter and eavesdropper are given by
\begin{gather}
\y_m = \hh_m \x + \n_m \quad \text{ and } \quad 
\y_e = \hh_e \x + \n_e, \label{eq:model1}
\end{gather}
respectively. Above, $\x$ denotes the $n_T \times 1$--dimensional transmitted signal vector. This channel input is subject to the following average power constraint:
\begin{gather}
\E\{\|\x\|^2\} = \tr(\K_x) \le P
\end{gather}
where $\tr$ denotes the trace operation and $\K_x = \E\{\x \x^\dd\}$ is the covariance matrix of the input. In (\ref{eq:model1}), $n_R \times 1$--dimensional $\y_m$ and $n_E \times 1$--dimensional $\y_e$ represent the received signal vectors at the legitimate receiver and eavesdropper, respectively. Moreover, $\n_m$ with dimension $n_R \times 1$ and $\n_e$ with dimension $n_E \times 1$ are independent, zero-mean Gaussian random vectors with $E\{\n_m \n_m^\dd\} = N_m \I$ and $E\{\n_e \n_e^\dd\} = N_e \I$, where $\I$ is the identity matrix. The signal-to-noise ratio is defined as
\begin{gather} \label{eq:snr}
\tsnr = \frac{\E\{\|\x\|^2\}}{\E\{\|\n_m\|^2\}} = \frac{P}{n_R N_m}.
\end{gather}
Finally, in the channel models, $\hh_m$ is the $n_R \times n_T$--dimensional channel matrix between the transmitter and legitimate receiver, and  $\hh_e$ is the $n_E \times n_T$--dimensional channel matrix between the transmitter and eavesdropper. While being fixed deterministic matrices in unfaded channels, $\hh_m$ and $\hh_e$ in fading channels are random matrices whose components denote the fading coefficients between the corresponding antennas at the transmitting and receiving ends.

\section{Secrecy in the Low-SNR Regime} \label{sec:lowsnrsecrecy}

Recently, in \cite{Khisti} and \cite{Oggier}, it has been shown that when the channel matrices $\hh_m$ and $\hh_e$ are fixed for the entire transmission period and are known to all three terminals\footnote{The assumption of perfect channel knowledge can, for instance, be justified in scenarios in which a base station, which knows the channels of the users, attempt to transmit confidential messages to a user and hence treat the other users as eavesdroppers.}, then the secrecy capacity in nats per dimension is given by\footnote{Unless stated otherwise, \!all \!logarithms throughout the paper are to the base $e$.}
\begin{align}\label{eq:secrecycap}
C_s = &\frac{1}{n_R}\max_{\substack{\K_x \succeq \0 \\ \tr(\K_x) \le P}}  \log \det \left(\I + \frac{1}{N_m} \hh_m \K_x \hh_m^\dd\right)
-\log \det \left(\I + \frac{1}{N_e} \hh_e \K_x \hh_e^\dd\right) \text{ nats/s/Hz/dimension}
\end{align}
where the maximization is over all possible input covariance matrices $\K_x \succeq \0$\footnote{$\succeq$ and $\succ$ denote positive semidefinite and positive definite partial orderings, respectively, for Hermitian matrices. If $\mathbf{A} \succeq \mathbf{B}$, then $\mathbf{A} - \mathbf{B}$ is a positive semidefinite matrix. Similarly, $\mathbf{A} \succ \mathbf{B}$ implies that $\mathbf{A} - \mathbf{B}$ is positive definite.} subject to a trace constraint. We note that since $\log \det \left(\I + 1/N_m \hh_m \K_x \hh_m^\dd \right)$ is a concave function of $\K_x$, the objective function in (\ref{eq:secrecycap}) is in general neither concave nor convex in $\K_x$, making the identification the optimal input covariance matrix a difficult task for arbitrary $\tsnr$ levels.

In this paper, we concentrate on the low-$\tsnr$ regime. In this regime, the behavior of the secrecy capacity can be accurately predicted by its first and second derivatives with respect to $\tsnr$ at $\tsnr = 0$:
\begin{align}
C_s(\tsnr) &= \dot{C}_s(0) \tsnr + \frac{\ddot{C}_s(0)}{2} \tsnr^2 + o(\tsnr^2). \label{eq:lowsnrapprox}
\end{align}
Moreover, $\dot{C}_s(0)$ and $\ddot{C}_s(0)$ also enable us to analyze the energy efficiency in the low-$\tsnr$ regime through the following notions \cite{Verdu}:
\begin{gather}\label{eq:ebno-so}
\frac{E_b}{N_0}_{s,\min} = \frac{\log 2}{\dc} \quad \text{and} \quad \so = \frac{2 \left[\dc\right]^2}{-\ddc}
\end{gather}
where $\frac{E_b}{N_0}_{s,\min}$ denotes the minimum bit energy required for reliable communication under secrecy constraints (or equivalently minimum energy per secret bit), and $\so$ denotes the wideband slope
which is the slope of the secrecy capacity in bits/dimension/(3 dB) at the point $\frac{E_b}{N_0}_{s,\min}$.
These quantities provide a linear
approximation of
the secrecy capacity in the low-$\tsnr$ regime. 
While $\frac{E_b}{N_0}_{s,\min}$ is a performance measure for vanishing $\tsnr$, $\so$ together with $\frac{E_b}{N_0}_{s,\min}$ characterize the performance at low but nonzero $\tsnr$s. We note that the formula for the minimum bit energy is valid if $C_s$ is a concave function of $\tsnr$, which we show later in the paper.

\subsection{First and Second Derivatives of the Secrecy Capacity}

Through the following result, we identify the first and second derivatives of the secrecy capacity at $\tsnr = 0$.

\begin{theo} \label{theo:secrecyderivatives}
The first derivative of the secrecy capacity in (\ref{eq:secrecycap}) with respect to $\tsnr$ at $\tsnr = 0$ is
\begin{gather}\label{eq:capfirstderiv}
\dot{C}_s(0) = [\lm(\bPhi)]^+ = \left\{
\begin{array}{ll}
\lm(\bPhi) & \text{if } \lm(\bPhi)>0
\\
0 & \text{else}
\end{array}\right.
\end{gather}
where
$
\bPhi = \hh_m^\dd \hh_m - \frac{N_m}{N_e}\hh_e^\dd \hh_e
$. $\dot{C}_s(0)$ can be achieved by choosing the input covariance matrix as $\K_x = P \, \bu \bu^\dd$ where $P$ denotes the average power and $\bu$ is the normalized eigenvector that corresponds to $\lm(\bPhi)$.

Moreover, the second derivative of the secrecy capacity at $\tsnr = 0$ is given by
\begin{align}
\ddot{C}_s(0) = -n_R \min_{\substack{\{\alpha_i\} \\ \alpha_i \in [0,1] \, \forall i \\ \sum_{i=1}^l \alpha_i = 1}} \sum_{i,j = 1}^l &\alpha_i \alpha_j \bigg( |\bu_j^\dd \hh_m^\dd \hh_m \bu_i|^2
- \frac{N_m^2}{N_e^2} |\bu_j^\dd \hh_e^\dd \hh_e \bu_i|^2\bigg) 1\{\lm(\bPhi > 0)\} \label{eq:capsecondderiv}
\end{align}
where $l$ is the multiplicity of $\lm(\bPhi)>0$, $\{\bu_i\}$ are the eigenvectors that span the maximal-eigenvalue eigenspace of $\bPhi$, and $1\{\lm(\bPhi) > 0 \} = \left\{
\begin{array}{ll}
1 & \text{if } \lm(\bPhi) > 0
\\
0 & \text{else}
\end{array}\right.
$ is the indicator function. The second derivative is achieved by choosing $\K_x = P \sum_{i = 1}^{l} \alpha_i \bu_i \bu_i^\dd$ where the values of $\{\alpha_i\}$ are determined by the optimization problem in (\ref{eq:capsecondderiv}).
\end{theo}

\emph{Proof}: See Appendix \ref{app:proofoftheo}.

\begin{rem} \label{rem:nosecrecy}
 In the absence of secrecy constraints, the first and second derivatives of the MIMO capacity at $\tsnr = 0$ are \cite{Verdu}
\vspace{-.4cm}
\begin{align}
\dot{C}(0) = \lm(\hh_m^\dd \hh_m) \text{ and }
\ddot{C}(0) = -\frac{n_R}{l}\lm^2(\hh_m^\dd \hh_m)
\end{align}
where $l$ is the multiplicity of $\lm(\hh_m^\dd \hh_m)$. Hence, the first and second derivatives are achieved by transmitting in the maximal-eigenvalue eigenspace of $\hh_m^\dd \hh_m$, the subspace in which the transmitter-receiver channel is the strongest. Due to the optimality of the water-filling power allocation method, power should be equally distributed in each orthogonal direction in this subspace in order for the second derivative to be achieved.
\end{rem}

\begin{rem}
We see from Theorem \ref{theo:secrecyderivatives} that when there are secrecy constraints, we should at low $\tsnr$s transmit in the direction in which the transmitter-receiver channel is strongest \emph{with respect to the transmitter-eavesdropper channel} normalized by the ratio of the noise variances. For instance, $\dot{C}_s(0)$ can be achieved by beamforming in the direction in which the eigenvalue of $\bPhi$ is maximized. On the other hand, if $\lm(\bPhi)$ has a multiplicity, the optimization problem in (\ref{eq:capsecondderiv}) should be solved to identify how power should be allocated to different orthogonal directions in the maximal-eigenvalue eigenspace so that the second-derivative $\ddot{C}_s(0)$ is attained. In general, the optimal power allocation strategy is neither water-filling nor beamforming. For instance, consider parallel Gaussian channels for both transmitter-receiver and transmitter-eavesdropper links, and assume that $\hh_m^\dd \hh_m = \text{diag}(5, 4, 2)$ and $\hh_e^\dd \hh_e = \text{diag}(2, 1, 1)$ where $\text{diag()}$ is used to denote a diagonal matrix with components provided in between the parentheses. Assume further that the noise variances are equal, i.e., $N_m = N_e$. Then, it can be easily seen that $\lm(\bPhi) = 3$ and has a multiplicity of $2$. Solving the optimization problem in (\ref{eq:capsecondderiv}) provides $\alpha_1 = 5/12$ and $\alpha_2 = 7/12$. Hence, approximately, $42\%$ of the power is allocated to the channel for which the transmitter-receiver link has a strength of $5$, and $58\%$ is allocated for the channel with strength $4$.
\end{rem}

\begin{rem}
When $\lm(\bPhi) > 0$ is distinct, then beamforming in the direction in which $\lambda(\bPhi)$ is maximized is optimal in the sense of achieving both $\dot{C}_s(0)$ and $\ddot{C}_s(0)$. Moreover, in this case, we have
\begin{gather}
\ddot{C}_s(0)  = -n_R \left( \|\hh_m \bu_1\|^4 - \frac{N_m^2}{N_e^2} \|\hh_e \bu_1\|^4\right)
\end{gather}
where $\bu_1$ is the eigenvector that corresponds to $\lm(\bPhi)$.
\end{rem}

\begin{rem} \label{rem:eigenvaluediff}
From \cite[Theorem 4.3.1]{matrixbook}, we know that for two Hermitian matrices $\mathbf{A}$ and $\mathbf{B}$ with the same dimensions, we have
\begin{gather}
\lm(\mathbf{A}+\mathbf{B}) \le \lm(\mathbf{A}) + \lm(\mathbf{B}).
\end{gather}
Applying this result to our setting yields
\begin{gather}
\lm(\bPhi) \le \lm(\hh_m^\dd \hh_m) -\lambda_{\min}\left(\frac{N_m}{N_e}\hh_e^\dd \hh_e \right).
\end{gather}
Therefore, we conclude from Remark \ref{rem:nosecrecy} that when $\lm(\bPhi) > 0$, secrecy constraints diminish the first derivative $\dot{C}_s(0)$ at least by a factor of $\lambda_{\min}\left(\frac{N_m}{N_e}\hh_e^\dd \hh_e \right)$ when compared to the case in which there are no such constraints.
\end{rem}

\begin{rem}
In the case in which the transmitter has a single antenna (i.e, $n_T = 1$), the channel matrices become column vectors. Denoting these column vectors as $\h_m$ and $\h_e$, we can immediately see from the result of Theorem \ref{theo:secrecyderivatives} that
\begin{align}
\dot{C}_s(0) &= \left[\|\h_m\|^2 - \frac{N_m}{N_e}\|\h_e\|^2\right]^+
\quad \text{and} \quad
\ddot{C}_s(0) = -n_R\left[ \|\h_m\|^4 - \frac{N_m^2}{N_e^2} \|\h_e\|^4\right]^+.
\end{align}
Similarly, if each terminal has a single antenna (i.e., $n_T = n_R = n_E = 1$), the results of Theorem \ref{theo:secrecyderivatives} specialize to
\begin{align}
\dot{C}_s(0) &= \left[|h_m|^2 - \frac{N_m}{N_e}|h_e|^2\right]^+
\quad \text{and} \quad
\ddot{C}_s(0) = -\left[ |h_m|^4 - \frac{N_m^2}{N_e^2} |h_e|^4\right]^+.
\end{align}
\end{rem}

Heretofore, we have considered the secrecy capacity which is obtained by finding the optimal input covariance matrix that maximizes the secrecy rate
\begin{align}
I_s(\tsnr)  = \frac{1}{n_R} \left[\log \det \left(\I + \frac{1}{N_m} \hh_m \K_x \hh_m^\dd\right)
-\log \det \left(\I + \frac{1}{N_e} \hh_e \K_x \hh_e^\dd\right)\right]^+. \label{eq:secrecyrate}
\end{align}
Hence, for a given input covariance matrix $\K_x$, the expression in (\ref{eq:secrecyrate}) provides the rate of secure communication. Using the same techniques as in the proof of Theorem \ref{theo:secrecyderivatives}, we can immediately obtain the following characterization.
\begin{cor} \label{cor:secrecyderivs}
For a given input covariance matrix $\K_x$, the first derivative of the secrecy rate in (\ref{eq:secrecyrate}) with respect to $\tsnr$ at $\tsnr = 0$ is
\begin{align}
\dot{I}_s(0) = \left[\tr\left( \hh_m \tK_x \hh_m^\dd  - \frac{N_m}{N_e} \hh_e \tK_x \hh_e^\dd \right)\right]^+ &= \left[\tr\left( \left(\hh_m^\dd \hh_m  - \frac{N_m}{N_e} \hh_e^\dd \hh_e\right) \tK_x  \right)\right]^+
\\
&=\left[\tr\left( \bPhi \tK_x  \right)\right]^+
\end{align}
where $\tK_x = \frac{1}{P}\K_x$ is the normalized input covariance matrix, and $\bPhi$ is again defined as $\bPhi = \hh_m^\dd \hh_m  - \frac{N_m}{N_e} \hh_e^\dd \hh_e $. The second derivative of the secrecy rate at $\tsnr = 0$ is given by
\begin{align}
\ddot{I}_s(0) = -n_R \, \tr\left( \left(\hh_m \tK_x \hh_m^\dd\right)^2  - \frac{N_m^2}{N_e^2} \left(\hh_e \tK_x \hh_e^\dd\right)^2 \right) \, 1\left\{ \tr\left( \bPhi \tK_x  \right) > 0\right\}. \label{eq:secondderivsecrecyrate}
\end{align}
\end{cor}

\emph{Proof}: See Appendix \ref{app:proofofcor}.

For instance, if the transmitter opts to uniformly allocate the power across the antennas, the covariance matrix becomes $\K_x = \frac{P}{n_T} \I$. Hence, we have $\tK_x = \frac{1}{n_T} \I$. In this case, we can readily see from Corollary \ref{cor:secrecyderivs} that we have
\begin{align}
\dot{I}_s(0) = \left[\frac{1}{n_T}\tr\left( \bPhi \right)\right]^+ = \left[\frac{1}{n_T} \sum_i \lambda_i(\bPhi) \right]^+ \le [\lm(\bPhi)]^+.
\end{align}
This result indicates that when we have uniform power allocation, the first derivative of the secrecy rate  is proportional to the average of the eigenvalues of $\bPhi$ rather than the maximum eigenvalue, and we in general experience, as expected, a loss in performance.

\begin{figure}
\begin{center}
\includegraphics[width=\figsize\textwidth]{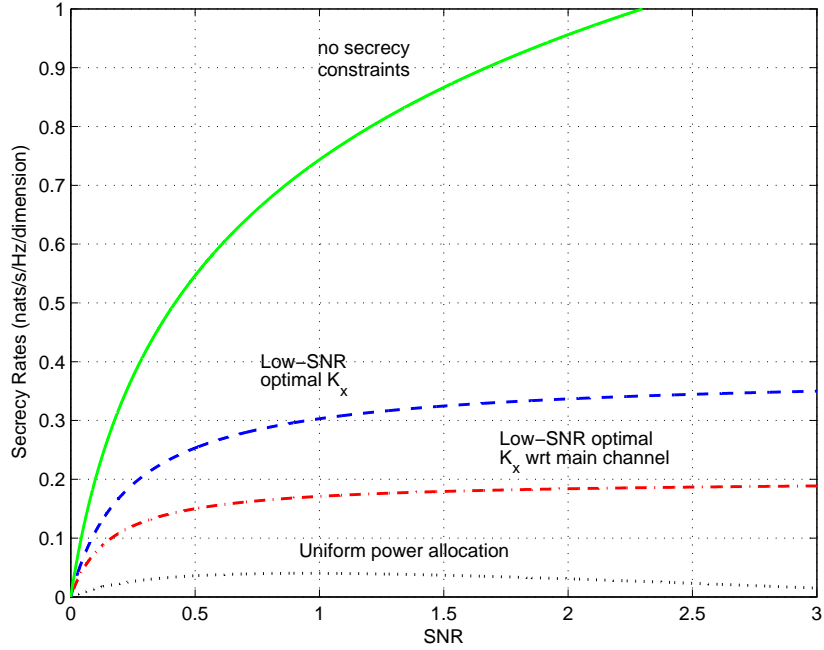}
\caption{Secrecy rates in nats/s/Hz/dimension vs. $\tsnr$. }\label{fig:secrecyrates}
\end{center}
\end{figure}
We now illustrate the theoretical results through numerical analysis. We consider a system in which all terminals have 3 antennas, i.e., $n_T = n_R = n_E = 3$. Assume that the channel matrices are
\begin{align}
\hh_m = \left[\begin{array}{ccc}
  1 & 0.8 & 0.5 \\
  0.3 & 1 & 0.1 \\
  0.1 & 0.2 & 0.1
\end{array}\right] \quad \text{and} \quad
\hh_e = \left[\begin{array}{ccc}
  0.5 & 0.4 & 1 \\
  0.7 & 0.1 & 0.5 \\
  0.3 & 0.5 & 0.1
\end{array}\right].
\end{align}
Assume further that $N_m = N_e = 1$. It can be easily verified that the maximum eigenvalue of the matrix $\bPhi = \hh_m^\dd \hh_m - \hh_e^\dd \hh_e$ is distinct and is equal to $\lm(\bPhi) = 1.6298$. The eigenvector that corresponds to $\lm(\bPhi)$ is $\bu^\dd = [-0.4677 \,\, -0.8823 \,\,\,\,\,\, 0.054]$. Therefore, the covariance matrix that is optimal in the sense of achieving both the first and second derivatives of the secrecy capacity is $\K_x = P \, \bu \bu^\dd$. In Figure \ref{fig:secrecyrates} in which secrecy rates are plotted as a function of $\tsnr$, the dashed curve shows the secrecy rates achieved when this input covariance matrix is employed. Note that this secrecy rate curve is optimally close to the secrecy capacity in the low-$\tsnr$ regime as it has the same first and second derivatives. Note also that for the considered model, we have $\dot{C}_s(0) = \lm(\bPhi) = 1.6298$. Fig. 1 also provides secrecy rates for two suboptimal choices of $\K_x$. The dot-dashed curve plots the secrecy rates when $K_x = P \, \bv \bv^\dd$ where $\bv$ is the eigenvector that corresponds to $\lm(\hh_m^\dd \hh_m)$. Hence, transmission in this case is performed in the direction in which the channel between the transmitter and legitimate receiver is strongest. Note that this strategy is optimal in the low-$\tsnr$ regime if there are no secrecy considerations. However, as we also observe in the figure, it is in general suboptimal in the wiretap channel model. Even the slope at zero $\tsnr$ is smaller. Indeed, the slope is $\dot{I}_s(0) = \tr(\bPhi \tK_x) = 1.2444$. In Fig. \ref{fig:secrecyrates}, we also plot the secrecy rates (with the dotted curve) when the power is uniformly allocated across the antennas. In this case, we have $\dot{I}_s(0) = \frac{1}{3}\, \tr(\bPhi) = 0.18$, which is about $11\%$ of $\dot{C}_s(0)$. Inefficiency of uniform power allocation is further evidenced in the observation that the secrecy rates start diminishing as $\tsnr$ is increased beyond 0.94, due to the fact that transmission is also possibly being conducted in the directions in which the eavesdropper's channel is strong and consequently, increasing the power improves the eavesdropper's ability to wiretap the channel. Finally, as a comparison, we plot in Fig. \ref{fig:secrecyrates} the rates achieved in the absence of secrecy constraints when $K_x = P \, \bv \bv^\dd$ with $\bv$ as defined above. For this case, the first derivative of the capacity is $\lm(\hh_m^\dd \hh) = 2.7676$.

\subsection{Minimum Energy per Secret Bit} \label{subsec:minenergy}

In this section, we study the energy required to send information both reliably and securely. In particular, we investigate the minimum energy required to send one secret bit. Before identifying the minimum energy per secret bit, we first show that the secrecy capacity is concave in $\tsnr$.

\begin{prop} \label{prop:concavity}
The secrecy capacity $C_s$ achieved under the average power constraint $\E\{\|\x\|^2\} \le P$ is a concave function of $\tsnr$.
\end{prop}

\emph{Proof:} Concavity can be easily shown using the time-sharing argument. Assume that at power level $P_1$ and signal-to-noise ratio $\tsnr_1$, the optimal input is $\x_1$, which satisfies $\E\{\|\x_1\|^2\} \le P_1$, and the secrecy capacity is $C_s(\tsnr_1)$. Similarly, for $P_2$ and $\tsnr_2$, the optimal input is $\x_2$, which satisfies $\E\{\|\x_2\|^2\} \le P_2$, and the secrecy capacity is $C_s(\tsnr_2)$. Now, we assume that the transmitter performs time-sharing by transmitting at two different power levels using $\x_1$ and $\x_2$. More specifically, in $\theta$ fraction of the time, the transmitter uses the input $\x_1$, transmits at most at $P_1$, and achieves the secrecy rate $C_s(\tsnr_1)$. In the remaining $(1-\theta)$ fraction of the time, the transmitter employs $\x_2$, transmits at most at $P_2$, and achieves the secrecy rate $C_s(\tsnr_2)$. Hence, this scheme overall achieves the average secrecy rate of
\begin{gather} \label{eq:timesharingrate}
\theta C_s(\tsnr_1) + (1-\theta) C_s(\tsnr_2)
\end{gather}
by transmitting at the level $\theta \E\{\|\x_1\|^2\} + (1-\theta) \E\{\|\x_2\|^2\}  \le P_\theta = \theta P_1 + (1-\theta) P_2$.  The average signal-to-noise ratio is $\tsnr_\theta = \theta \tsnr_1 + (1-\theta) \tsnr_2$. Therefore, the secrecy rate in (\ref{eq:timesharingrate}) is an achievable secrecy rate at $\tsnr_\theta$. Since the secrecy capacity is the maximum achievable secrecy rate, the secrecy capacity at $\tsnr_\theta$ is larger than that in (\ref{eq:timesharingrate}), i.e.,
\begin{align}
C_s(\tsnr_\theta) &= C_s(\theta \tsnr_1 + (1-\theta) \tsnr_2)
\ge \theta C_s(\tsnr_1) + (1-\theta) C_s(\tsnr_2),
\end{align}
showing the concavity. \hfill $\blacksquare$

We further note that the concavity can also be shown using the following facts. As also discussed in \cite{Liu-Shamai}, MIMO secrecy capacity is obtained by proving in the converse argument that the considered upper bound is tight and
\begin{gather}
C_s = \max_{p(\x)} \min_{p(\y_r^{'}, \y_e^{'} | \x) \in \mathcal{D} } I(\x ; \y_r^{'} | \y_e^{'})
\end{gather}
where $\mathcal{D}$ is the set of joint conditional density functions $p(\y_r^{'}, \y_e^{'} | \x)$ that satisfy $p(\y_r^{'}| \x) = p(\y_r | \x)$ and $p(\y_e^{'}| \x) = p(\y_e | \x)$. Note that for fixed channel distributions, the mutual information $I(\x ; \y_r^{'} | \y_e^{'})$ is a concave function of the input distribution $p(\x)$. Since the pointwise infimum of a set of concave functions is concave \cite{convex}, $f(p(\x)) = \min_{p(\y_r^{'}, \y_e^{'} | \x) \in \mathcal{D} } I(\x ; \y_r^{'} | \y_e^{'})$ is also a concave function of $p(\x)$. Concavity of the functional $f$ and the fact that maximization is over input distributions satisfying $\E\{\|\x\|^2\} \le P$ lead to the concavity of the secrecy capacity with respect to $\tsnr$.

The energy per secret bit normalized by the noise variance at the legitimate receiver is defined as
\begin{gather}
\frac{E_b}{N_0}_{s} = \frac{\tsnr}{C_s(\tsnr)} \log 2.
\end{gather}
As mentioned before, since the secrecy capacity is a concave function of $\tsnr$, the minimum energy per secret bit is achieved as $\tsnr \to 0$ and hence is given by
\begin{gather}
\frac{E_b}{N_0}_{s,\min} = \lim_{\tsnr \to 0} \frac{\tsnr}{C_s(\tsnr)} \log 2= \frac{\log 2}{\dc}.
\end{gather}
We can now write the following corollary to Proposition \ref{prop:concavity} and Theorem \ref{theo:secrecyderivatives}.

\begin{cor}
The minimum bit energy attained under secrecy constraints (i.e., minimum energy per secret bit) is
\begin{gather}
\frac{E_b}{N_0}_{s,\min} =  \frac{\log 2} {[\lm(\bPhi)]^+}.
\end{gather}
\end{cor}

\begin{rem}
From Remark \ref{rem:eigenvaluediff}, we can write for $\lm(\bPhi) > 0$
\begin{align}
\frac{E_b}{N_0}_{s,\min} = \frac{\log 2} {\lm(\bPhi)} &\ge \frac{\log 2} {\lm(\hh_m^\dd \hh_m) -\lambda_{\min}\left(\frac{N_m}{N_e}\hh_e^\dd \hh_e \right)}\ge \frac{\log 2} {\lm(\hh_m^\dd \hh_m)} = \frac{E_b}{N_0}_{\min} \label{eq:ebnomin_nosecrecy}
\end{align}
where $\frac{E_b}{N_0}_{\min}$ in (\ref{eq:ebnomin_nosecrecy}) denotes the minimum bit energy in the absence of secrecy constraints. Hence, in general, secrecy requirements increase the energy expenditure. When secure communication is not possible, $[\lm(\bPhi)]^+ = 0$ and $\frac{E_b}{N_0}_{s,\min} = \infty$.
\end{rem}

\begin{rem}
Energy costs of secrecy can easily be identified in the case in which the transmitter has a single-antenna. Clearly, the minimum bit energy in the presence of secrecy is strictly greater than that in the absence of such constraints:
\begin{gather}
\frac{E_b}{N_0}_{s,\min} = \frac{\log 2}{\left[\|\h_m\|^2 - \frac{N_m}{N_e} \|\h_e\|^2\right]^+} > \frac{\log 2}{\|\h_m\|^2} = \frac{E_b}{N_0}_{\min}
\end{gather}
when  $\frac{N_m}{N_e} \|\h_e\|^2 > 0 $. Furthermore, the energy requirement increases monotonically as the value of $\frac{N_m}{N_e} \|\h_e\|^2$ increases. Indeed, when $\frac{N_m}{N_e} \|\h_e\|^2 = \|\h_m\|^2$, secure communication is not possible and $\frac{E_b}{N_0}_{s,\min} = \infty$.
\end{rem}

The expression for the wideband slope $S_0$ can be readily obtained by plugging in the expressions in (\ref{eq:capfirstderiv}) and (\ref{eq:capsecondderiv}) into that in (\ref{eq:ebno-so}):
\begin{align} \nonumber
\so = \frac{2 \left[\dc\right]^2}{-\ddc} = \frac{2 \left([\lm(\bPhi)]^+\right)^2}{n_R \min_{\substack{\{\alpha_i\} \\ \alpha_i \in [0,1] \, \forall i \\ \sum_{i=1}^l \alpha_i = 1}} \sum_{i,j = 1}^l \alpha_i \alpha_j \bigg( |\bu_j^\dd \hh_m^\dd \hh_m \bu_i|^2
- \frac{N_m^2}{N_e^2} |\bu_j^\dd \hh_e^\dd \hh_e \bu_i|^2\bigg) 1\{\lm(\bPhi > 0)\}}.
\end{align}

\begin{figure}
\begin{center}
\includegraphics[width=\figsize\textwidth]{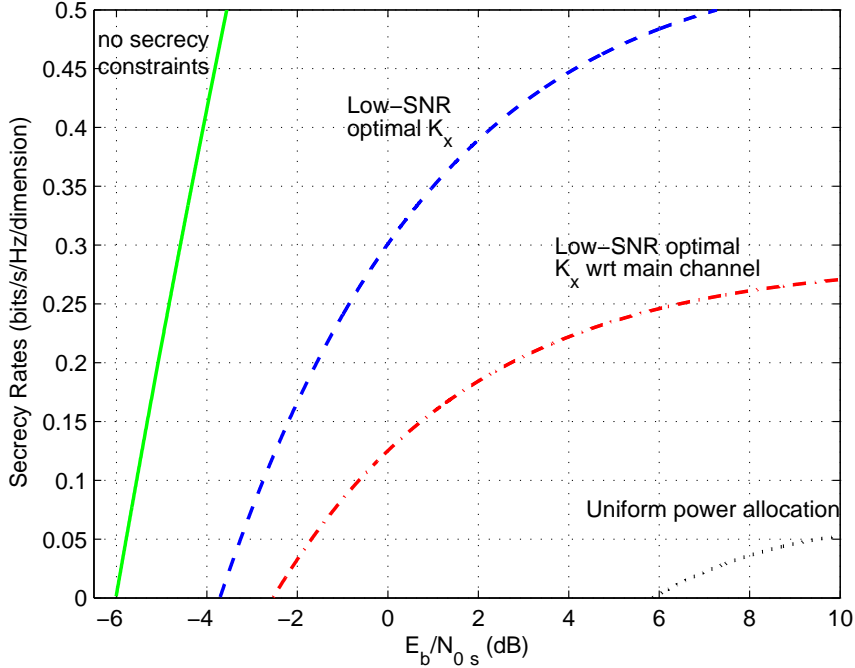}
\caption{Secrecy rates in bits/s/Hz/dimension vs. energy per secret bit $\frac{E_b}{N_0}_s$. }\label{fig:secretbitenergy}
\end{center}
\end{figure}
In Fig. \ref{fig:secretbitenergy}, we plot the secrecy rates in bits/s/Hz/dimension as a function of the energy per secret bit $\frac{E_b}{N_0}_s$ under the same assumptions and channel model as in Fig. \ref{fig:secrecyrates}. We see, as predicted, that the minimum bit energy is attained in all cases as $\tsnr$ and hence rates approach zero. While the minimum bit energy is $\frac{E_b}{N_0}_{\min} = \frac{\log 2}{\lm (\hh_m^\dd \hh_m)} = -6.01$ dB in the absence of secrecy constraints, the minimum bit energy per secret bit is $\frac{E_b}{N_0}_{s,\min} = \frac{\log 2}{\lm (\bPhi)} = -3.71$ dB. Therefore, secrecy constraints lead to an increase of 2.3 dB in the minimum energy requirements. We also note that the energy cost of secrecy increases as secrecy rates increase. Moreover, we observe that the suboptimal choices of $\K_x$ induce additional energy penalties. When we have $\K_x = P \bv \bv^\dd$ where $\bv$ is the eigenvector that corresponds to $\lm(\hh_m^\dd \hh_m)$, the minimum bit energy is $-2.54$ dB. In the case of uniform power allocation, the minimum bit energy requirement jumps to $5.85$ dB.

\section{The Impact of Fading} \label{sec:fading}

In this section, we assume that the channel matrices $\hh_m$ and $\hh_e$ are random matrices whose components are ergodic random variables, modeling fading in wireless transmissions. We again assume that realizations of these matrices are perfectly known by all the terminals. As discussed in \cite{Liang}, fading channel can be regarded as a set of parallel subchannels each of which corresponds to a particular fading realization. Hence, in each subchannel, the channel matrices are fixed similarly as in the channel model considered in the previous section. In \cite{Liang}, Liang \emph{et al.} have shown that having independent inputs for each subchannel is optimal and the secrecy capacity of the set of parallel subchannels is equal to the sum of the capacities of subchannels. Therefore, the secrecy capacity of fading channels can be be found by averaging the secrecy capacities attained for different fading realizations.

We assume that the transmitter is subject to a short-term power constraint. Hence, for each channel realization, the same amount of power is used and we have $\tr(\K_x) \le P$. With this assumption, the transmitter is allowed to perform power adaptation in space across the antennas, but not across time. Under such constraints, it can easily be seen from the above discussion that the average secrecy capacity in fading channels is given by
\begin{align}\label{eq:avgsecrecycap}
C_s = &\frac{1}{n_R} \, \E_{\hh_m, \hh_e} \Bigg\{\max_{\substack{\K_x \succeq \0 \\ \tr(\K_x) \le P}}  \log \det \left(\I + \frac{1}{N_m} \hh_m \K_x \hh_m^\dd\right)\nonumber
-\log \det \left(\I + \frac{1}{N_e} \hh_e \K_x \hh_e^\dd\right)\Bigg\}
\end{align}
where the expectation is with respect to the joint distribution of $(\hh_m, \hh_e)$. Note that the only difference between (\ref{eq:secrecycap}) and (\ref{eq:avgsecrecycap}) is the presence of expectation in (\ref{eq:avgsecrecycap}). Due to this similarity, the following result can be obtained immediately as a corollary to Theorem \ref{theo:secrecyderivatives}.
\begin{cor}
The first derivative of the average secrecy capacity in (\ref{eq:avgsecrecycap}) with respect to $\tsnr$ at $\tsnr = 0$ is
\begin{gather}
\dot{C}_s(0) = \E_{\hh_m, \hh_e}\{[\lm(\bPhi)]^+\}
\end{gather}
where again
$
\bPhi = \hh_m^\dd \hh_m - \frac{N_m}{N_e}\hh_e^\dd \hh_e.
$
The second derivative of the average secrecy capacity at $\tsnr = 0$ is given by
\begin{align} \label{eq:avgcapsecondderiv}
\ddot{C}_s(0) = -n_R \E_{\hh_m, \hh_e} &\Bigg\{\min_{\substack{\{\alpha_i\} \\ \alpha_i \in [0,1] \, \forall i \\ \sum_{i=1}^l \alpha_i = 1}} \!\!\sum_{i,j = 1}^l \alpha_i \alpha_j \bigg( |\bu_j^\dd \hh_m^\dd \hh_m \bu_i|^2  - \frac{N_m^2}{N_e^2} |\bu_j^\dd \hh_e^\dd \hh_e \bu_i|^2\bigg)  1\{\lm(\bPhi) > 0\} \Bigg\}
\end{align}
where $1\{\cdot\}$ again denotes the indicator function, $l$ is the multiplicity of $\lm(\bPhi)>0$, and $\{\bu_i\}$ are the eigenvectors that span the maximal-eigenvalue eigenspace for particular realizations of $\hh_m$ and $\hh_e$.
\end{cor}

\begin{rem}
Similarly as in the unfaded case, $\dot{C}_s(0)$ is achieved by always transmitting in the maximal-eigenvalue eigenspace of the realizations of the matrix $\bPhi$. In order to achieve the second derivative, optimal values of $\{\alpha_i\}$ (or equivalently the optimal power allocation across the antennas) should be identified again for each possible realization of $\bPhi$.
\end{rem}

\begin{rem}
In the case in which $n_T = 1$, the first and second derivatives of the average secrecy capacity become 
\begin{align}
\dot{C}_s(0) &= \E_{\h_m, \h_e}\left\{\left[\|\h_m\|^2 - \frac{N_m}{N_e}\|\h_e\|^2\right]^+\right\}
\quad \text{and} \quad
\ddot{C}_s(0) &= -n_R\E_{\h_m, \h_e}\left\{\left[\|\h_m\|^4 - \frac{N_m^2}{N_e^2}\|\h_e\|^4\right]^+\right\}. \nonumber
\end{align}
\end{rem}

Similarly as in Section \ref{subsec:minenergy}, we can identify the minimum energy per secret bit as follows.

\begin{cor}
The minimum energy per secret bit required in fading channels is
\begin{gather}
\frac{E_b}{N_0}_{s,\min} =  \frac{\log 2} {\E_{\hh_m, \hh_e}\{[\lm(\bPhi)]^+\}}.
\end{gather}
\end{cor}

\begin{rem}
Fading has a potential to improve the low-$\tsnr$ performance and hence the energy efficiency. To illustrate this, we consider the following example. Assume $n_T = n_R = n_E = 1$. Consider first the unfaded Gaussian channel in which the deterministic channel coefficients are $h_m = h_e = 1$. For this case, we have \vspace{-.3cm}
\begin{align}
\dc = \left[1 - \frac{N_m}{N_e}\right]^+ \text{ and } \frac{E_b}{N_0}_{s,\min} = \frac{\log 2}{\left[1 - \frac{N_m}{N_e}\right]^+}.
\end{align}
Now, consider a Rayleigh fading environment and assume that $h_m$ and $h_e$ are independent, zero-mean, circularly symmetric Gaussian random variables with variances $E\{|h_m|^2\} = E\{|h_e|^2\} = 1$. Then, we can easily find that \vspace{-.3cm}
\begin{gather}
\dot{C}_s(0) = \E_{h_m, h_e}\left\{\left[|h_m|^2 - \frac{N_m}{N_e}|h_e|^2\right]^+\right\} = \frac{N_e}{N_m + N_e}
\end{gather}
leading to $\frac{E_b}{N_0}_{s,\min} = \frac{\log 2}{\frac{N_e}{N_m + N_e}}$.
Note that if $N_e > 0$,
$
\frac{N_e}{N_m + N_e} > \left[1 - \frac{N_m}{N_e}\right]^+.
$ Hence, fading strictly improves the low-$\tsnr$ performance by increasing $\dc$ and decreasing the minimum bit energy even without performing power control over time. Further gains are possible with power adaptation. Another interesting observation is the following. In unfaded channels, if $N_m \ge N_e$, the minimum bit energy is infinite and secure communication is not possible. On the other hand, in fading channels, the bit energy is finite as long as $N_m$ is finite and $N_e > 0$. Clearly, even if $N_m \ge N_e$, favorable fading conditions enable secure transmission in fading channels.
The positive impact of fading on secrecy rates especially at low $\tsnr$s has been discussed for instance in \cite{Bloch} and \cite{Gopala}. Here, we provide a similar observation from the energy efficiency perspective.
\end{rem}

\begin{figure}
\begin{center}
\includegraphics[width=\figsize\textwidth]{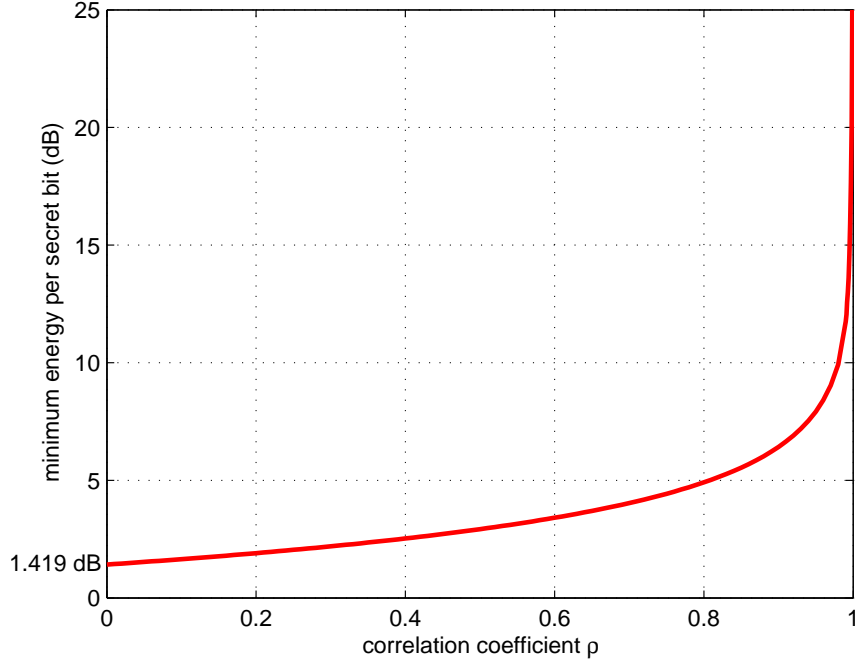}
\caption{Minimum energy per secret bit $\frac{E_b}{N_0}_{\min,s}$ vs. correlation coefficient $\rho$. }\label{fig:minenergy}
\end{center}
\end{figure}

Above, we have assumed that the fading coefficients $h_m$ and $h_e$ are independent. Next, we demonstrate that the gains are still observed even if the channel coefficients are correlated. We again assume that $h_m$ and $h_e$ are zero-mean, circularly symmetric Gaussian random variables with $\E\{|h_m|^2\} = \E\{|h_e|^2\} = 1$. Let us denote $r_m = |h_m|, z_m = |h_m|^2$ and $r_e = |h_e|, z_e = |h_e|^2$. Using the following bivariate Rayleigh probability density function
given in \cite[Equation 6.2]{commbook}
\begin{gather}
f(r_m, r_e) = \frac{4r_m r_e}{1-\rho} \, e^{-\frac{1}{1-\rho}(r_m^2 + r_e^2)} \, I_0\left( \frac{2\sqrt{\rho \, r_m^2 r_e^2}}{1-\rho}\right),
\end{gather}
we can easily obtain the bivariate exponential density as
\begin{gather}
f(z_m, z_e) = \frac{1}{1-\rho} \, e^{-\frac{1}{1-\rho}(z_m + z_e)} \, I_0\left( \frac{2\sqrt{\rho \, z_m z_e}}{1-\rho}\right).
\end{gather}
In the above formulation, $I_0$ denotes the zeroth order modified Bessel function of the first kind. Moreover, $\rho$ denotes the power correlation coefficient, which is related to the correlation coefficients of the underlying Gaussian random variables $h_m$ and $h_e$, and is given by \cite{commbook}
\begin{gather}
\rho = \frac{\left|\E\{h_m h_e^*\}\right|^2}{\E\{|h_m|^2\} E\{|h_e|^2\}}
\end{gather}
under the assumption that $h_m$ and $h_e$ are zero-mean. With this characterization, we can now easily compute $\dot{C}_s(0) = \E_{h_m, h_e}\left\{\left[|h_m|^2 - \frac{N_m}{N_e}|h_e|^2\right]^+\right\}$, from which we can obtain the minimum energy per secret bit $\frac{E_b}{N_0}_{s,\min} = \frac{\log 2}{\dc}$. In Fig. \ref{fig:minenergy}, the minimum energy per secret bit is plotted as a function of the correlation coefficient $\rho$. When $\rho = 0$ and hence the channel coefficients are independent, we have $\frac{E_b}{N_0}_{s,\min} = \frac{\log 2}{\frac{N_e}{N_m + N_e}} = \frac{\log2}{0.5} = 1.419$ dB. As the correlation increases, the minimum bit energy value increases. However, note that the bit energy values are finite unless there is full correlation. Note further that if there were no fading, we would have $\frac{E_b}{N_0}_{s,\min} = \frac{\log 2}{\left[1 - \frac{N_m}{N_e}\right]^+} = \infty$ (recalling the assumption that $N_m = N_e = 1$). Hence, in general, correlated fading provides improvements in secure communication as well.

\begin{figure}
\begin{center}
\includegraphics[width=\figsize\textwidth]{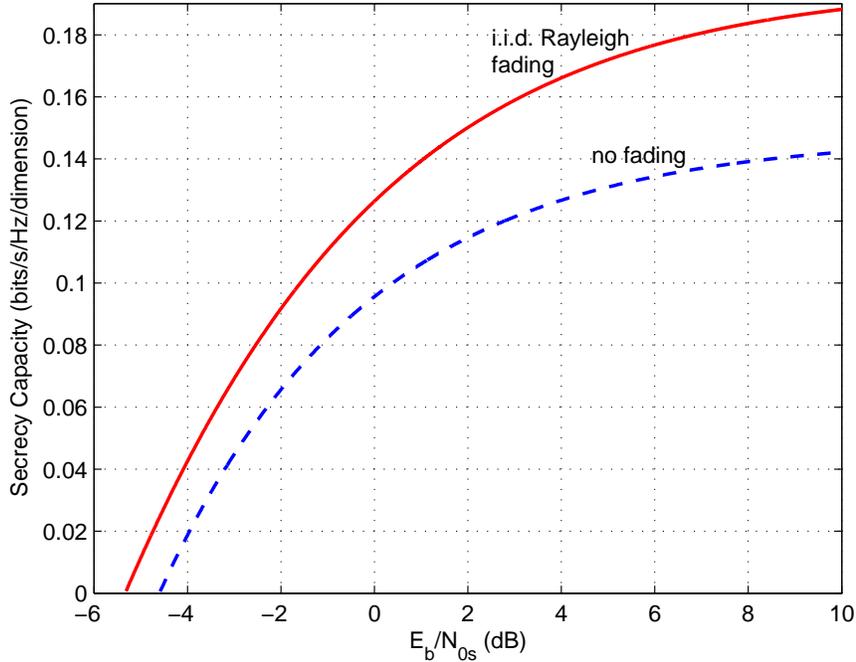}
\caption{Secrecy capacity in bits/s/Hz/dimension vs. energy per secret bit $\frac{E_b}{N_0}_s$ when $n_T = 1, n_R = 5,$ and $n_E = 3$.}\label{fig:fadingnR5nE3}
\end{center}
\end{figure}

Above, improvements in the minimum energy per secret bit, which is attained as $\tsnr$ vanishes, are discussed. In general, fading is beneficial in terms of energy efficiency at nonzero $\tsnr$ levels as well. This is demonstrated in Fig. \ref{fig:fadingnR5nE3}. In this figure, we plot the secrecy capacity when $n_T = 1, n_R = 5$, and $n_E = 3$. We consider two scenarios: no fading and i.i.d. Rayleigh fading. In the case in which there is no fading, we assume that the channel coefficients are all equal to 1. In the fading scenario, we assume that the channel vectors $\h_m$ and $\h_e$ consist of independent and identically distributed, zero-mean Gaussian components each with unit variance, i.e., $\E\{|h_{m,i}|^2\} = 1$ and $\E\{|h_{e,i}|^2\} = 1$ for all $i$. We additionally assume that $\h_m$ and $\h_e$ are independent of each other. Note that under these assumptions, $\|\h_m\|^2$ and $\|\h_e\|^2$ are independent chi-square random variables with $2n_R$ and $2n_E$ degrees of freedom, respectively. In Fig. \ref{fig:fadingnR5nE3}, we observe that better performance is achieved in the presence of fading. As readily seen, the minimum energy per secret bit required in Rayleigh fading is smaller. Moreover, for a given secrecy capacity value, less bit energy is needed in the presence of fading. Indeed, energy gains tend to increase at higher values of secrecy capacity. For instance, when $C_s = 0.14$ bits/s/Hz/dimension, we have a gain of approximately $8$ dB in $\frac{E_b}{N_0}_s$. Note that this is a substantial improvement in energy efficiency.

\begin{figure}
\begin{center}
\includegraphics[width=\figsize\textwidth]{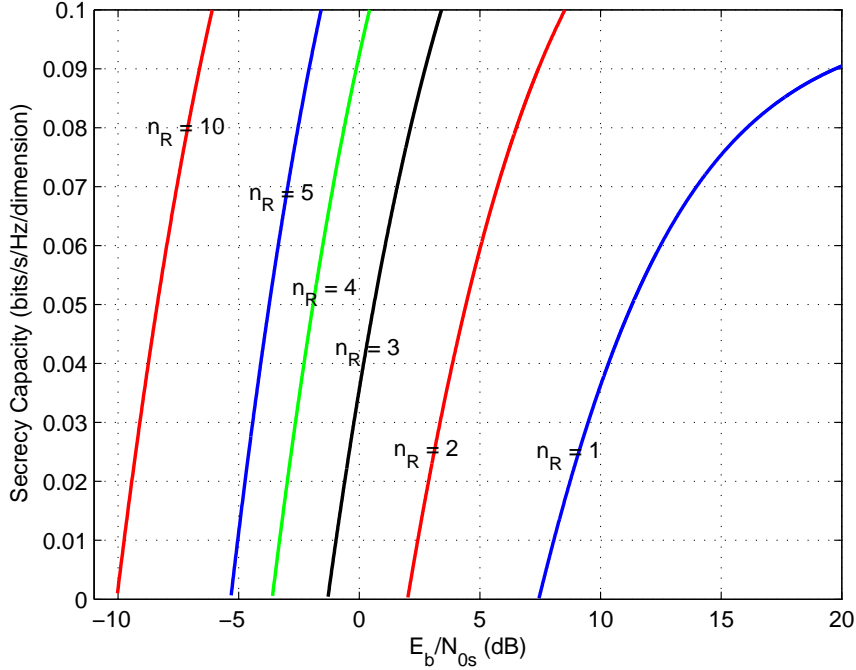}
\caption{Secrecy capacity in bits/s/Hz/dimension vs. energy per secret bit $\frac{E_b}{N_0}_s$ in i.i.d. Rayleigh fading when $n_T = 1$ and $n_E = 3$.}\label{fig:fadingnE3}
\end{center}
\end{figure}

As another benefit, fading enables secure communication, which otherwise is not possible in a non-fading environment. For instance, under the assumptions that $n_T = 1$ and all channel coefficients are equal to 1, secrecy capacity is zero if the legitimate receiver has the same as or less number of antennas than the eavesdropper. However, this is not necessarily the case in fading scenarios. Due to the randomness of fading coefficients, there are instants with non-zero probabilities, in which the main channel is stronger than the eavesdropper's channel even though $n_R \le n_E$. This is illustrated in Fig. \ref{fig:fadingnE3} in which we plot the secrecy capacity as a function of $\frac{E_b}{N_0}_s$ in i.i.d. Rayleigh fading for different values of $n_R$ when we have $n_T = 1$ and $n_E = 3$. Note that even when $n_R \le 3$, we require finite bit energy for secure communications. In the above-mentioned non-fading scenario, we would have $\frac{E_b}{N_0}_s = \infty$. Additionally, we note that performance, as expected, improves and less energy per secret bit is required as the number of receive antennas $n_R$ increases.

\section{Conclusion} \label{sec:conclusion}

In this paper, we have analyzed the MIMO secrecy capacity in the low-$\tsnr$ regime. We have obtained expressions for the first and second derivatives of the secrecy capacity at $\tsnr = 0$. Using these expressions, we have identified the optimal transmission strategies in the low-$\tsnr$ regime under secrecy constraints. In particular, we have shown that it is optimal to transmit in the maximal-eigenvalue eigenspace of the matrix $\bPhi = \hh_m^\dd \hh_m - \frac{N_m}{N_e}\hh_e^\dd \hh_e$. We have compared the low-$\tsnr$ results with those obtained in the absence of secrecy constraints, and quantified the degradation in the performance. We have determined the minimum bit energy required for secure and reliable communications in the presence of an eavesdropper. We have shown that secrecy in general increases the bit energy requirements. We have also noted that the suboptimal choices of transmission strategies can incur additional energy penalties. Numerical results are provided to illustrate the theoretical findings. Following the analysis for the fixed channel, we have investigated the low-$\tsnr$ secrecy capacity in the presence of fading. We have generalized our derivative results to apply to the perfectly-known fading channel. We have demonstrated the benefits of fading in terms of energy efficiency.

\appendix

\subsection{Proof of Theorem \ref{theo:secrecyderivatives}} \label{app:proofoftheo}

We first note that the input covariance matrix $\K_x = \E\{\x \x^\dd\}$ is by definition a positive semidefinite Hermitian matrix. As a Hermitian matrix, $\K_x$ can be written as \cite[Theorem 4.1.5]{matrixbook}
\begin{align} \label{eq:spectral}
\K_x = \U \bLambda \U^\dd
\end{align}
where $\U$ is a unitary matrix and $\bLambda$ is a real diagonal matrix. Using (\ref{eq:spectral}), we can also express $\K_x$ as
\begin{gather}
\K_x = \sum_{i = 1}^{n_T} d_i \bu_i \bu_i^\dd
\end{gather}
where $\{d_i\}$ are the diagonal components of $\bLambda$, and $\{\bu_i\}$ are the column vectors of $\U$ and form an orthonormal set. Assuming that the input uses all the available power, we have $\tr(\K_x) = \sum_{i = 1}^{n_T}{d_i} = P$. Noting that $\K_x$ is positive semidefinite and hence  $d_i \ge 0$, we can write $d_i = \alpha_i P$ where $\alpha_i \in [0,1] $ $\forall i$ and $\sum_{i = 1}^{n_T} \alpha_i = 1$. Now, the secrecy rate achieved with a particular covariance matrix $\K_x$ can be expressed as
\begin{align}
I_s(\tsnr) &= \frac{1}{n_R} \Bigg( \log \det \left(\I + n_R \, \tsnr \sum_{i = 1}^{n_T} \alpha_i \hh_m \bu_i \bu_i^\dd \hh_m^\dd\right) -\log \det \left(\I + \frac{n_R N_m}{N_e} \, \tsnr \sum_{i = 1}^{n_T} \alpha_i \hh_e \bu_i \bu_i^\dd \hh_e^\dd\right) \Bigg) \nonumber.
\end{align}
where $\tsnr$ is defined in (\ref{eq:snr}). As also noted in \cite{Verdu}, we can easily show that
\vspace{-.5cm}
\begin{align}
\frac{d}{dv} \log \det (\I + v \mathbf{A}) \bigg|_{v = 0} &= \tr(\mathbf{A}), \label{eq:derivlogdet}
\\
\frac{d^2}{dv^2} \log \det (\I + v \mathbf{A}) \bigg|_{v = 0} &= - \tr(\mathbf{A}^2). \label{eq:deriv2logdet}
\end{align}
Now, using (\ref{eq:derivlogdet}), we obtain the following expression for the first derivative of the secrecy rate $I_s$ with respect to $\tsnr$ at $\tsnr = 0$:
\begin{align}
\hspace{-.2cm}\dot{I}_s(0) &= \sum_{i = 1}^{n_T} \alpha_i \left(\tr(\hh_m \bu_i \bu_i^\dd \hh_m^\dd) - \frac{N_m}{N_e}\tr(\hh_e \bu_i \bu_i^\dd \hh_e^\dd) \right)
\\
&= \sum_{i = 1}^{n_T} \alpha_i \left(\bu_i^\dd \hh_m^\dd \hh_m \bu_i  - \frac{N_m}{N_e}\bu_i^\dd \hh_e^\dd \hh_e \bu_i  \right) \label{eq:dotI2}
\\
&= \sum_{i = 1}^{n_T} \alpha_i \bu_i^\dd \left(\hh_m^\dd \hh_m - \frac{N_m}{N_e}\hh_e^\dd \hh_e\right) \bu_i
= \sum_{i = 1}^{n_T} \alpha_i \bu_i^\dd  \bPhi \bu_i \label{eq:dotI4}
\end{align}
where (\ref{eq:dotI2}) follows from the property that $\tr(\mathbf{A}\mathbf{B}) = \tr(\mathbf{B}\mathbf{A})$. Also, in (\ref{eq:dotI4}), we have defined $\bPhi = \hh_m^\dd \hh_m - \frac{N_m}{N_e}\hh_e^\dd \hh_e$. Since $\bPhi$ is a Hermitian matrix and $\{\bu_i\}$ are unit vectors, we have \cite[Theorem 4.2.2]{matrixbook}
\begin{gather} \label{eq:upperbound-lambda}
\bu_i^\dd  \bPhi \bu_i \le \lambda_{\max}(\bPhi) \quad \forall i
\end{gather}
where $\lambda_{\max}(\bPhi)$ denotes the maximum eigenvalue of the matrix $\bPhi$. Recall that $\alpha_i \in [0,1]$ and $\sum_{i} \alpha_i = 1$. Then, from (\ref{eq:upperbound-lambda}), we obtain
\vspace{-.3cm}
\begin{align}
\dot{I}_s(0) = \sum_{i = 1}^{n_T} \alpha_i \bu_i^\dd  \bPhi \bu_i \le \lambda_{\max}(\bPhi). \label{eq:upperbounddotI}
\end{align}
Note that this upper bound can be achieved if, for instance, $\alpha_1 = 1$ and $\alpha_i = 0$ $\forall i \neq 1$, and $\bu_1$ is chosen as the eigenvector that corresponds to the maximum eigenvalue of $\bPhi$. Heretofore, we have implicitly assumed that $\lm(\bPhi) > 0$ and all the available power is used to transmit the information in the direction of the maximum eigenvalue. If $\lm(\bPhi) \le 0$, then all eigenvalues of $\bPhi$ are less than or equal to zero, and hence $\bPhi$ is a negative semidefinite matrix. In this situation, none of the channels of the legitimate receiver is stronger than those corresponding ones of the eavesdropper. In such a case, secrecy capacity is zero. Therefore, if $\lm(\bPhi) \le 0$, we have $\dot{C}_s(0) = 0$. Finally, we conclude from (\ref{eq:upperbounddotI}) and the above discussion that the first derivative of the secrecy capacity with respect to $\tsnr$ at $\tsnr = 0$ is given by
\begin{align}
\dot{C}_s(0) = [\lm(\bPhi)]^+ = \left\{
\begin{array}{ll}
\lm(\bPhi) & \text{if } \lm(\bPhi)>0
\\
0 & \text{else}
\end{array}\right..
\end{align}
If $\lm(\bPhi) > 0$ is distinct, $\dot{C}_s(0)$ is achieved when we choose $\K_x = P \bu_1 \bu_1^\dd$ where $\bu_1$ is the eigenvector that corresponds to $\lm(\bPhi)$. Therefore, beamforming in the direction in which the eigenvalue of $\bPhi$ is maximized is optimal in the sense of achieving the first derivative of the secrecy capacity in the low-$\tsnr$ regime. More generally, if $\lm(\bPhi) > 0 $ has a multiplicity, any covariance matrix in the following form achieves the first derivative: \vspace{-.3cm}
\begin{gather} \label{eq:covariance-opt}
\K_x = P \sum_{i = 1}^{l} \alpha_i \bu_i \bu_i^\dd
\end{gather}
where $l$ is the multiplicity of the maximum eigenvalue, $\{\bu_i\}_{i = 1}^l$ are the eigenvectors that span the maximal-eigenvalue eigenspace of $\bPhi$, and $\{\alpha_i\}_{i = 1}^l$ are constants, taking values in $[0,1]$ and having the sum $\sum_{i=1}^l \alpha_i = 1$. Therefore, transmission in the maximal-eigenvalue eigenspace is necessary to achieve $\dot{C}_s(0)$.

Next, we consider the second derivative of the secrecy capacity. Again, when $\lm(\bPhi) \le 0$, the secrecy capacity is zero and therefore $\ddc = 0$. Hence, in the following, we consider the case in which $\lm(\bPhi) > 0$. Suppose that the input covariance matrix is chosen as in (\ref{eq:covariance-opt}) with a particular set of $\{\alpha_i\}$. Then, using (\ref{eq:deriv2logdet}), we can obtain
\begin{align}
\ddot{I_s}(0) &= -n_R \,\, \tr \left( \left( \sum_{i = 1}^{l} \alpha_i \hh_m \bu_i  \bu_i^\dd \hh_m^\dd \right)^2\right) + n_R \frac{N_m^2}{N_e^2} \,\, \tr\left( \left( \sum_{i = 1}^{l} \alpha_i \hh_e \bu_i  \bu_i^\dd \hh_e^\dd \right)^2\right)
\\
&= -n_R \sum_{i,j} \alpha_i \alpha_j \left( |\bu_j^\dd \hh_m^\dd \hh_m \bu_i|^2 - \frac{N_m^2}{N_e^2} |\bu_j^\dd \hh_e^\dd \hh_e \bu_i|^2\right) \label{eq:ratesecondderiv}
\end{align}
where  (\ref{eq:ratesecondderiv}) is obtained by using the fact that $\tr(\mathbf{A}\mathbf{B}) = \tr(\mathbf{B}\mathbf{A})$ and performing some straightforward manipulations. Note again that $\{\bu_i\}$ are the eigenvectors spanning the maximal-eigenvalue eigenspace of $\bPhi$. Being necessary to achieve the first derivative, the covariance structure given in (\ref{eq:covariance-opt}) is also necessary to achieve the second derivative. Therefore, the second derivative of the secrecy capacity at $\tsnr=0$ is the maximum of the expression in (\ref{eq:ratesecondderiv}) over all possible values of $\{\alpha_i\}$. Hence,
\begin{align}
\ddot{C}_s(0) = -n_R \!\!\!\!\min_{\substack{\{\alpha_i\} \\ \alpha_i \in [0,1] \, \forall i \\ \sum_{i=1}^l \alpha_i = 1}} \sum_{i,j} &\alpha_i \alpha_j \bigg( |\bu_j^\dd \hh_m^\dd \hh_m \bu_i|^2 - \frac{N_m^2}{N_e^2} |\bu_j^\dd \hh_e^\dd \hh_e  \bu_i|^2\bigg) \label{eq:capsecondderivintheproof}
\end{align}
Since $\ddc$ is equal to the expression in (\ref{eq:capsecondderivintheproof}) when $\lm(\bPhi) > 0$ and is zero otherwise,  the final expression in (\ref{eq:capsecondderiv}) is obtained by multiplying the formula in (\ref{eq:capsecondderivintheproof}) with the indicator function $1\{\lm(\bPhi) > 0\}$. \hfill $\blacksquare$

\subsection{Proof of Corollary \ref{cor:secrecyderivs}} \label{app:proofofcor}

The secrecy rate is expressed as
\begin{align}
I_s(\tsnr)  = \frac{1}{n_R} \left[\log \det \left(\I + \frac{1}{N_m} \hh_m \K_x \hh_m^\dd\right)
-\log \det \left(\I + \frac{1}{N_e} \hh_e \K_x \hh_e^\dd\right)\right]^+. \label{eq:secrecyrateproof}
\end{align}
Defining the normalized input covariance matrix as $\tK_x = \frac{1}{P} \K_x$, we can rewrite the secrecy rate as
\begin{align}
I_s(\tsnr)  = \frac{1}{n_R} \left[\log \det \left(\I + n_R \, \tsnr \, \hh_m \tK_x \hh_m^\dd\right)
-\log \det \left(\I + \frac{n_R N_m}{N_e} \tsnr \, \hh_e \tK_x \hh_e^\dd\right)\right]^+. \label{eq:secrecyrateproof2}
\end{align}
where we, similarly as before, have $\tsnr = \frac{P}{n_R N_m}$. Then, using (\ref{eq:derivlogdet}), we immediately have \begin{gather}
\dot{I}_s(0) = \left[\tr\left( \hh_m \tK_x \hh_m^\dd \right) - \tr\left(\frac{N_m}{N_e} \hh_e \tK_x \hh_e^\dd\right)\right]^+
= \left[\tr\left( \hh_m \tK_x \hh_m^\dd  - \frac{N_m}{N_e} \hh_e \tK_x \hh_e^\dd\right)\right]^+.
\end{gather}
In order to obtain the second derivative, we can apply (\ref{eq:deriv2logdet}) to the positive part of the secrecy rate to obtain
\begin{align}
\ddot{I}_s(0) &= -n_R \left( \tr\left( \left(\hh_m \tK_x \hh_m^\dd\right)^2 \right) - \tr\left(\frac{N_m^2}{N_e^2} \, \left(\hh_e \tK_x \hh_e^\dd\right)^2\right) \right)
\\
&= -n_R \left(\tr\left( \left(\hh_m \tK_x \hh_m^\dd\right)^2  - \frac{N_m^2}{N_e^2} \left(\hh_e \tK_x \hh_e^\dd\right)^2\right)\right).
\end{align}
Note that the above expression is the second derivative of the positive part of the secrecy rate, and hence applies only when the secrecy rate is positive. If the secrecy rate is zero, the second derivative is also zero, and hence we have the indicator function in the final expression in (\ref{eq:secondderivsecrecyrate}). \hfill $\blacksquare$

\end{spacing}
\end{document}